\documentclass[aps,prb, twocolumn,longbibliography,superscriptaddress]{revtex4-1}
\usepackage[T1]{fontenc}
\usepackage{latexsym}
\usepackage{graphicx} 
\usepackage{epstopdf}
\usepackage{amsmath}  
\usepackage{amssymb}
\usepackage{comment}
\usepackage{cases}
\usepackage{lipsum}
\usepackage{float}
\usepackage{tikz}
\usepackage{makecell}
\usepackage{soul}
\usepackage{color}
\usepackage[breaklinks=true]{hyperref}

\DeclareMathOperator*{\SumInt}{%
\mathchoice%
  {\ooalign{$\displaystyle\sum$\cr\hidewidth$\displaystyle\int$\hidewidth\cr}}
  {\ooalign{\raisebox{.14\height}{\scalebox{.7}{$\textstyle\sum$}}\cr\hidewidth$\textstyle\int$\hidewidth\cr}}
  {\ooalign{\raisebox{.2\height}{\scalebox{.6}{$\scriptstyle\sum$}}\cr$\scriptstyle\int$\cr}}
  {\ooalign{\raisebox{.2\height}{\scalebox{.6}{$\scriptstyle\sum$}}\cr$\scriptstyle\int$\cr}}
}

\begin{document}

\title{Robust spin and charge excitations throughout high-$T_c$-cuprate phase diagram from incipient Mottness}
\author{M. Fidrysiak}
\email{maciej.fidrysiak@uj.edu.pl}
\affiliation{Institute of Theoretical Physics, Jagiellonian University, ul. {\L}ojasiewicza 11, 30-348 Krak{\'o}w, Poland }
\author{J. Spa{\l}ek}%
\email{jozef.spalek@uj.edu.pl}
\affiliation{Institute of Theoretical Physics, Jagiellonian University, ul. {\L}ojasiewicza 11, 30-348 Krak{\'o}w, Poland }

\begin{abstract}  
  The generic phase diagram of lightly hole-doped high-$T_c$-cuprates hosts antiferromagnetic insulating phase with well-defined spin-wave excitations. Contrary to the weak-coupling prediction, these modes persist up to the overdoped metallic regime as intense and dispersive paramagnons. Here we report on our study of the low-energy magnetic and charge excitations within the extended Hubbard model at strong-coupling, using a modified $1/N$ expansion method with a variational state serving as the saddle point solution. Despite clear separation of magnetic and Hubbard-$U$ energy scales, we find that incipient Mottness affects qualitatively dispersions and widths of magnetic modes throughout entire phase diagram. The obtained magnetic and charge dynamical structure factors agree semi-quantitatively with recent resonant $X$-ray and neutron scattering data for $\mathrm{La_{2-\mathit{x}}Sr_\mathit{x}CuO_4}$ and $\mathrm{(Bi,Pb)_2(Sr,La)_2CuO_{6+\delta}}$ at all available doping levels. The weak-coupling random-phase-approximation fails already for underdoped samples, pointing to the non-trivial intertwining of distinct energy scales in cuprate superconductors. The existence of a discrete charge mode which splits off the electron-hole continuum is also predicted.
\end{abstract}

\maketitle

\section{Introduction}

Formulation of fully testable theory of high-temperature superconductivity (HTSC) in the cuprates remains one of the most challenging problems in condensed matter physics. The strongly correlated nature of electrons involved in HTSC, demonstrated in numerous experiments,\cite{HufnerRepProgPhys2008,YoshidaJPSJ2012,KeimerNauture2015,ProustAnnualReview2019}  provides a firm reference point for theoretical modeling and has been under elaboration for the last two decades.\cite{OgataRepProgPhys2008,RanderiaBook2011,KaczmarczykNewJPhys2014,AlloulComptesRendusPhys2014}  The principal features of the phase diagram and other \emph{equilibrium} properties have been reproduced with a degree of success, albeit the pseudogap appearance and associated with it possibility of quantum-critical-point emergence has no unequivocal interpretation as yet.\cite{HusseyRepProgPhys2018,SpalekPhysRevB2017,FidrysiakJPhysCondensMatter2018} 

A new  impetus in the field has been provided by recent developments \cite{AmentPhysRevLett2009,HaverkortPhysRevLett2010,BraicovichPhysRevLett2010} in resonant inelastic $X$-ray scattering (RIXS), granting access to the detailed structure of spin and charge \cite{IshiiPhysRevB2017,HeptingNature2018,IshiiJPhysSocJapan2019,FumagalliPhysRevB2019,Lin_arXiV_2019} excitations in highly-doped materials. The major experimental finding, consistent among variety of cuprates, is the persistence of intense and dispersive antinodal paramagnons ranging from antiferromagnetic (AF) insulator to overdoped paramagnetic (PM) metal, and their rapid overdamping along the nodal direction.\cite{DeanNatMater2013,IshiiNatCommun2014,LeeNatPhys2014,GuariseNatCommun2014,WakimotoPhysRevB2015,MinolaPhysRevLett2017,IvashkoPhysRevB2017,PengNatPhys2017,MeyersPhysRevB2017,ChaixPhysRevB2018,Robarts_arXiV_2019}  Such a selective robustness against damping due to electron-hole (e-h) scattering is inconsistent with weak-coupling prediction and has sparked a discussion about the role those modes may play in HTSC.\cite{LeTaconNatPhys2011,JiaNatCommun2014,PengPhysRevB2018}  Analogous results have been reported for iron pnictides \cite{ZhouNatCommun2013} and iridates,\cite{GretarssonPhysRevLett2016}  pointing towards universality of their presence in correlated materials. Previous theoretical works \cite{ScalapinoRevModPhys2012,ChubukovChapter2003}  involved, among others, spin excitations as a viable mechanism of pairing in HTSC. Elucidating the microscopic mechanism underlying this behavior remains a challenge to theory. A successful interpretation of those \emph{dynamic} excitations would also provide a convincing strongly correlated picture of cuprate and related superconductors.

In our approach, the description is divided into the equilibrium \textit{static} part, driven by a combined effect of local correlations and exchange interactions,\cite{AndersonScience2007,LeeRevModPhys2006,JedrakPhsRevB2011,SpalekPhysRevB2017}  and \emph{dynamic} collective fluctuations around this reference static saddle-point solution. The purpose of this work is to offer the description of the latter dynamic excitations. Namely,  we show, starting from the Hubbard-model, that local electronic correlations qualitatively reorganize the magnetic excitation spectrum in hole-doped systems when compared to that resulting from the weak-coupling spin-fluctuation theory. Using a single set of doping independent model parameters for given cuprate family, we reproduce semi-quantitatively experimental magnetic-mode energies and anisotropic paramagnon damping for $\mathrm{La_{2-\mathit{x}}Sr_\mathit{x}CuO_4}$ and $\mathrm{(Bi,Pb)_2(Sr,La)_2CuO_{6+\delta}}$ throughout the phase diagram. To capture the interplay between collective-mode fluctuations and correlated itinerant electrons, a dedicated computational technique extending the variational wave function approach is developed. As a reference point,  we discuss the weak-coupling random phase approximation (RPA) results and demonstrate its inadequacy to doped high-$T_c$ cuprates. 

The technical aspects of the method formulation, paramagnon spectrum extraction procedure, phase stability analysis, and numerical details are provided in Appendices~\ref{appendix:method}-\ref{appendix:fitting}.

\begin{figure*}
  \onecolumngrid
  \centering
  \includegraphics[width=\linewidth]{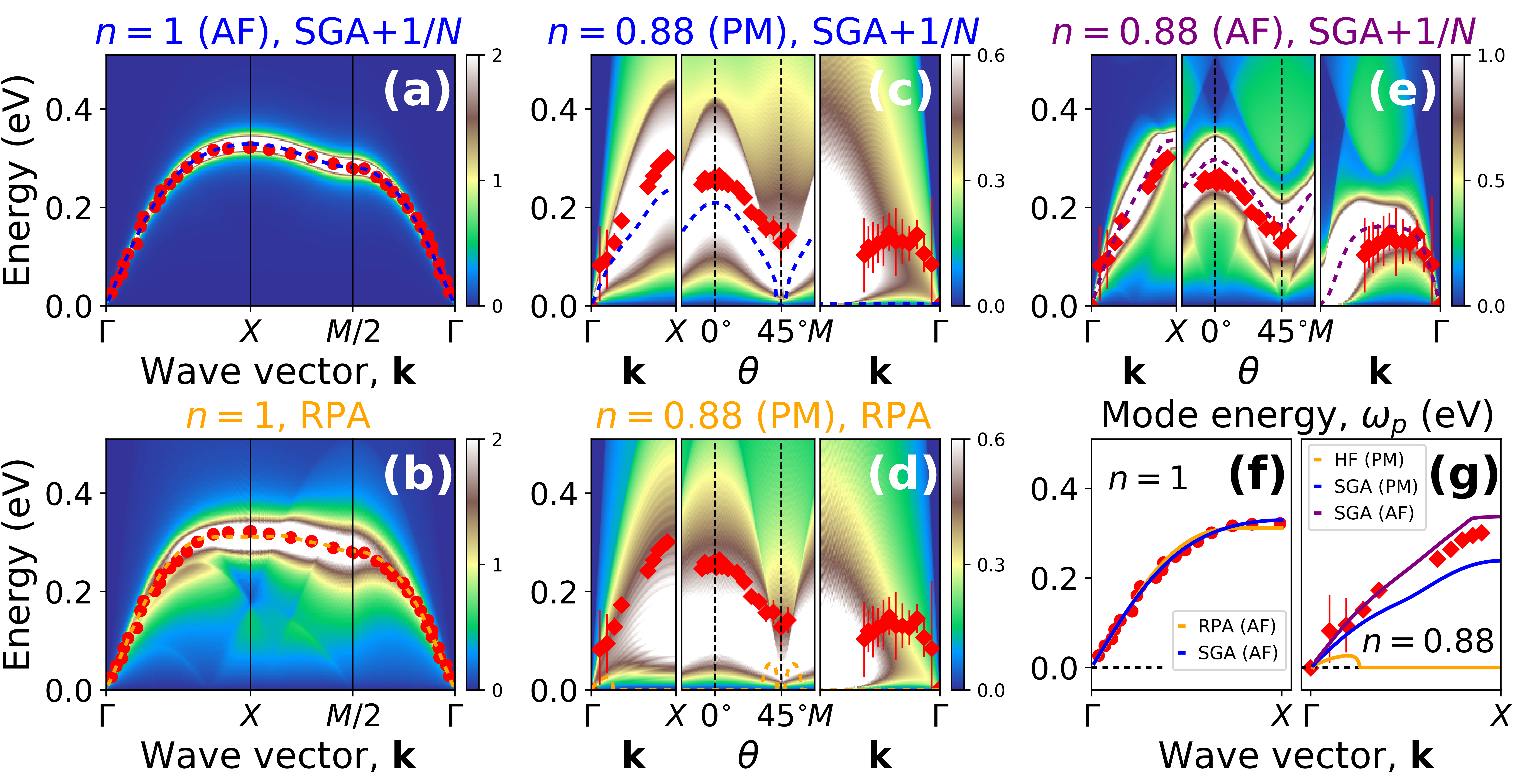}
  \caption{Calculated quantum spin-fluctuation spectra for undoped AF insulator [$n = 1$; panels (a-b)],  paramagnetic metal [$n = 0.88$; panels (c-d)], and antiferromagnetic metal [$n = 0.88$; panel (e)]. The model parameters are provided in Table~\ref{tab:parameters}. Dashed lines in (a-e) represent real parts of the quasiparticle pole $\omega_p$ (the so-called propagation frequency), obtained by damped-harmonic-oscillator modeling of the numerical data as detailed in Appendix~\ref{appendix:fitting}.  Experimental energies for $\mathrm{La_2CuO_4}$ [INS; actually shifted by $\mathbf{q}_\mathrm{AF} = (0.5, 0.5)$] and $\mathrm{La_{1.88}Sr_{0.12}CuO_4}$ (RIXS) are taken from \citenum{HeadingsPhysRevB2010} and \citenum{IvashkoPhysRevB2017}, respectively. Panels (f) and (g) provide the direct comparison of the calculated and experimental dispersions for $n=1.0$ and $n=0.88$, respectively. Angle $\theta$ in panels (c-e) parameterizes the Brillouin-zone arc $\mathbf{k}(\theta) \equiv 0.37 \cdot (\cos \theta, \sin \theta)$.}
  \label{fig:lsco}
  \twocolumngrid
\end{figure*}

\section{Model and method}

We start with a general one-band $t$-$J$-$U$ Hamiltonian $\hat{\mathcal{H}} = \sum_{\sigma, i \neq j} t_{ij} \hat{c}^{\dagger}_{i\sigma} \hat{c}_{i\sigma} + U \sum_i \hat{n}_{i\uparrow} \hat{n}_{i\downarrow} + J \sum_{\langle ij\rangle}\hat{\mathbf{S}}_i \hat{\mathbf{S}}_j$, retaining nearest- and next-nearest hopping integrals $t < 0$ and $t' = 0.25 |t|$, respectively. This model encompasses both $t$-$J$ and Hubbard model limits (see Ref.~\citenum{ZhangPhysRevLett2003}). Here $\hat{n}_{i\sigma} \equiv \hat{c}^\dagger_{i\sigma} \hat{c}_{i\sigma}$, $\hat{\mathbf{S}}_i$ is spin operator, and $U$ and $J$ are on-site repulsion and intersite exchange, respectively.  Spin- and charge dynamical susceptibilities are computed using modified $1/N$ expansion ($N$ counts fermionic flavors), so that the saddle point ($N=\infty$) coincides with the variational statistically-consistent (SGA) solution.\cite{JedrakPhsRevB2011,SpalekPhysRevB2017,FidrysiakJPhysCondensMatter2018}  This allows to capture both the Mott physics and long-wavelength collective excitations already at the leading expansion order, carried out with the help of analysis of the Ginzburg-Landau type, but of microscopic character. The present technique is formulated in terms of Grassmann variables ($\bar{\eta}_{i\sigma}$ and $\eta_{i\sigma}$), describing itinerant electrons, Grassmann bilinears $\hat{P}_i^{(\sigma, \sigma^\prime, \delta i, \mathrm{Re})} \propto [\bar{\eta}_{i+\delta i \sigma} \eta_{i\sigma^\prime} + \bar{\eta}_{i\sigma^\prime} \eta_{i+\delta i, \sigma}]/2$, $\hat{P}_i^{(\sigma, \sigma^\prime, \delta i, \mathrm{Im})} \propto -i [\bar{\eta}_{i+\delta i,\sigma} \eta_{i\sigma^\prime} - \bar{\eta}_{i\sigma^\prime} \eta_{i+\delta i,\sigma}]/2$, and corresponding composite fields ${P}_i^{(\sigma, \sigma^\prime, \delta i, \mathrm{Re}/\mathrm{Im})} \propto \langle \hat{P}_i^{(\sigma, \sigma^\prime, \delta i, \mathrm{Re}/\mathrm{Im})} \rangle$ accommodating collective modes (their complex counterparts are $P_{ij}^{\sigma\sigma^\prime} \equiv P_j^{\sigma\sigma^\prime i-j, \mathrm{Re}} + i P_j^{\sigma\sigma^\prime i-j,\mathrm{Im}}$). For example, $P_{ii}^{\sigma\sigma}(\tau) = n_{i\sigma}(\tau)$ and $P^{\uparrow\downarrow}_{ii} = S^{+}_i(\tau) \equiv m_i(\tau)$. Here $i$ and $\sigma$ are lattice- and spin indices, respectively.  We consider the imaginary-time action

\begin{align}
  \label{eq:action}
  \mathcal{S} = & \SumInt\limits_i d\tau  \boldsymbol{\eta}_i^\dagger (\partial_\tau - \mu) \boldsymbol{\eta}_i + \int d\tau E_G(\mathbf{P}_i, \{x\})  \nonumber\\ & + i \SumInt\limits_{ij} d\tau \boldsymbol{\xi}_{i}^T (\hat{\mathbf{P}}_{i}  - \mathbf{P}_{i}),
\end{align}

\noindent
where the vector notation indicates internal index summations and $E_G \equiv \langle \hat{\mathcal{H}} \rangle_G \equiv \langle\Psi_G| \hat{\mathcal{H}} | \Psi_G\rangle/\langle\Psi_G| \Psi_G\rangle$ is the SGA energy, evaluated using Gutzwiller-type wave function $|\Psi_G(\{x\})\rangle$ \cite{JedrakPhsRevB2011,SpalekPhysRevB2017,FidrysiakJPhysCondensMatter2018} ($\{x\}$ is the set of correlator parameters). By application of Wick's theorem, $E_G$ becomes a functional of $\mathbf{P}_i(\tau)$ and  $\{x\}$. The last term in Eq.~\eqref{eq:action} is the constraint enforcing that the composite fields follow the Hamiltonian dynamics. Finally, $E_G$ is expanded in fields $\mathbf{P}_i$ up to bilinear terms, and relevant correlation functions are evaluated to the leading non-trivial $1/N$ expansion order using the action $\mathcal{S}$. Susceptibility matrix is obtained by Fourier-transforming the expressions ${\boldsymbol{\chi}}_{ij}(\tau, \tau') = \int \mathbf{P}_i(\tau) \mathbf{P}_j^T(\tau') \exp(-\mathcal{S})$, analytic continuation to real frequencies and unfolding them to the PM Brillouin zone. The current scheme is free of Fierz ambiguity, making standard $1/N$ phase diagrams strongly-dependent on unphysical parameters.\cite{JaeckelPhysRevD2003,BaierPhysRevB2004}  This feature of the SGA+$1/N$ constitutes an essential improvement as it allows for an unbiased comparison with experiment. Extended discussion of the SGA+$1/N$ technique is provided in Appendix~\ref{appendix:method}. Although the method is general, here we analyze the simpler version of the $t$-$J$-$U$ model, namely the limit of the Hubbard model.

The generalized susceptibility $\boldsymbol{\chi}_{ij}(\tau, \tau^\prime)$ describes both the charge and spin dynamic susceptibilities. Those are calculated within the linear response theory starting from the saddle-point (SGA) approximation for the charge and spin correlation functions. The latter, in turn, are determined from the system of equations, which in the spin sector has the form

\begin{widetext}
  \begin{align}
    \label{eq:SS}
    \langle{\hat{\mathbf{S}}_i \hat{\mathbf{S}}_j}\rangle_G =&  g_{ ij}^{00} +  \sum\limits_{\sigma\sigma'} g_{ ij}^{n_\sigma n_{\sigma^\prime}} \langle\hat{n}_{i\sigma} \hat{n}_{j\sigma'}\rangle_0^c + g_{ij}^{mm^{*}} \langle\hat{m}_{i} \hat{m}_{j}^{\dagger}\rangle_0^c + g_{ ij}^{m^{*}m} \langle\hat{m}_{i}^{\dagger} \hat{m}_{j}\rangle_0^c + g_{ij}^{mm} \langle\hat{m}_{i} \hat{m}_{j}\rangle_0^c + g_{ ij}^{m^{*}m^{*}} \langle\hat{m}_{i}^{\dagger} \hat{m}_{j}^{\dagger}\rangle_0^c +\nonumber \\ & \sum\limits_\sigma g_{ij}^{n_\sigma m} \langle\hat{n}_{i\sigma} \hat{m}_{j}\rangle_0^c + \sum\limits_\sigma g_{ij}^{n_\sigma m^{*}} \langle\hat{n}_{i\sigma} \hat{m}_{j}^{\dagger}\rangle_0^c + \sum\limits_\sigma g_{ij}^{mn_\sigma} \langle \hat{m}_{i} \hat{n}_{j\sigma} \rangle_0^c + \sum\limits_\sigma g_{ij}^{m^{*}n_\sigma} \langle \hat{m}_{i}^{\dagger} \hat{n}_{j\sigma} \rangle_0^c,
\end{align}
\end{widetext}

\noindent
where the superscript ``c'' means that only connected diagrams should be retained. The charge correlation function $\langle \hat{n}_i \hat{n}_j\rangle_G$ has the same structure, but with different set of ``$g$'' coefficients. In those expressions, $g^{ab}_{ij}$ are polynomials of 12 parameters $\{ \lambda_{i/j,\sigma\sigma^\prime} \}$, $\lambda_{i/j,0}$, and $\lambda_{i/j,d}$, and which, in turn, are expressed via densities $\{ n_{i/j,\sigma}, m_{i/j} \}$  and variational parameters $\{ x_{i/j}\}$. Due to hermiticity of the correlator $\hat{P}_G$ in the variational wave function,\cite{SpalekPhysRevB2017,FidrysiakJPhysCondensMatter2018,JedrakPhsRevB2011}  the number of parameters for both magnetic and Coulomb repulsion parts reduces from 17 to 13. One sees explicitly the coupling between charge and spin degrees of freedom, as the terms are of comparable magnitude. In the Hartree-Fock (HF) approximation, composing the RPA approximation, most of the renormalization factors reduce trivially to either 1 or 0, and the coupling between charge and spin excitations disappears. In effect, we can see explicitly the difference between the two approaches (HF and SGA+$1/N$) and the results will be discussed explicitly below. Also, to make the analysis more transparent, we discuss here only the results for the Hubbard model as well as confront them in a quantitative manner with experiment.

To emphasize, the essence of our approach relies on converting the ground-state energy functional $E_G\{x\}$ into the form with the Grassmann variables and thus composing a new effective Hamiltonian, i.e., $\hat{\mathcal{H}}\{\eta_i, x\}$. In effect, the action~\eqref{eq:action} takes a standard finite-temperature field-theoretical many-particle form.\cite{NegeleBook1988}  To account for the correlations, the statistically consistent renormalized mean-field theory \cite{JedrakPhsRevB2011,SpalekPhysRevB2017,FidrysiakJPhysCondensMatter2018} is taken from the start, and the quantum fluctuations are included subsequently.

The model parameters adopted for $\mathrm{La_{2-\mathit{x}}Sr_\mathit{x}CuO_4}$ (LSCO) and $\mathrm{(Bi,Pb)_2(Sr,La)_2CuO_{6+\delta}}$ (Bi2201) are summarized in Table~\ref{tab:parameters}. Temperature $T$ is set so as to stay clear of spin-spiral states present in both HF and Gutzwiller phase diagrams.\cite{IgoshevJPCM2015}  This step allows to study  PM state in the  doping range inaccessible within previous $T=0$ operator Gutzwiller-method extensions.\cite{SeiboldPhysRevLett2001,SeiboldPhysRevB2006}  Contrary to SGA+$1/N$, application of RPA to the present situation requires violating hierarchy $k_B T \ll |t| \ll U$ (or, alternatively, taking unphysically small $U \approx 1.5 |t|$ \cite{GuariseNatCommun2014}), making it inadequate for the cuprates. Nevertheless, we present the RPA results for comparison. Numerical and phase-stability aspects are detailed in Appendices \ref{appendix:numerical_details} and \ref{appendix:phase_stability}.

\begin{figure}
  \centering
  \includegraphics[width=\linewidth]{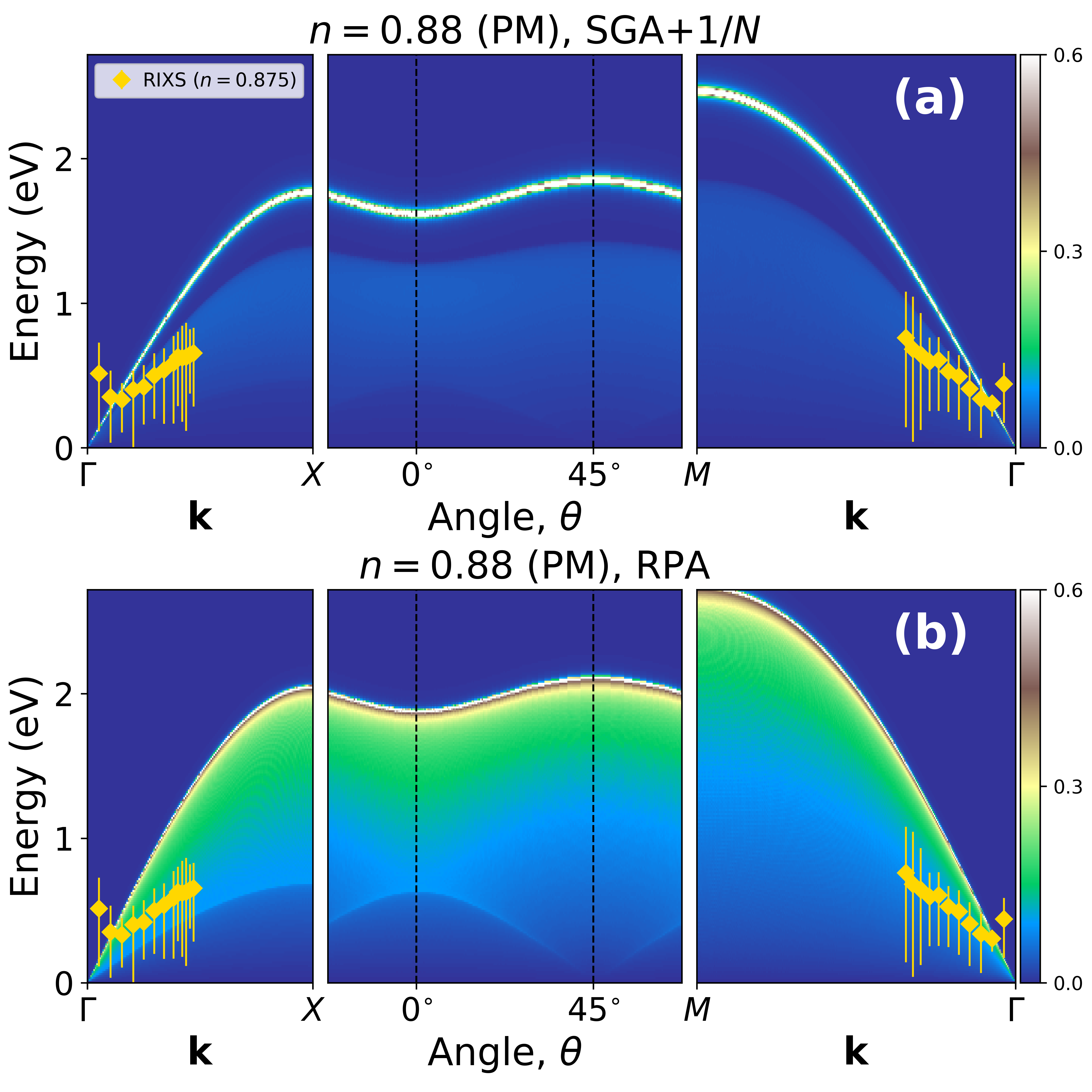}
  \caption{Calculated imaginary part of the dynamical charge susceptibility, $|t| \chi_C^{\prime\prime}(\mathbf{k}, \omega)$, along selected Brillouin-zone contours for LSCO at $n = 0.88$ (cf. Table~\ref{tab:parameters} for parameters). Panel (a) shows discrete charge-excitation mode splitting of the electron-hole continuum as obtained from SGA+$1/N$ approach. Analogous results obtained from RPA (b) are displayed for comparison. In the latter situation the mode appears at the upper threshold of the continuum. Angle $\theta$ parameterizes the arc $\mathbf{k}(\theta) \equiv 0.37 \cdot (\cos \theta, \sin \theta)$. Diamonds are RIXS data for $\mathrm{La_{2-\mathit{x}}(Br, Sr)_\mathit{x}CuO_4}$ at $n=0.875$.\cite{IshiiPhysRevB2017} }
  \label{fig:splitting}
\end{figure}

\section{Results}

In Fig.~\ref{fig:lsco} we display the calculated imaginary parts of transverse dynamical spin susceptibility along representative Brillouin-zone contours for $\mathrm{La_{2-\mathit{x}}Sr_\mathit{x}CuO_4}$ (see Table~\ref{tab:parameters} for model parameters). Red symbols are the inelastic neutron scattering \cite{HeadingsPhysRevB2010} and RIXS data for $\mathrm{La_{2-\mathit{x}}Sr_\mathit{x}CuO_4}$,\cite{IvashkoPhysRevB2017}  respectively. In panels (a-b) we display the spin-wave spectrum for half-filled ($n=1$) AF insulator, calculated using the present SGA+$1/N$ (a) and RPA (b) techniques, both of which accurately reproduce the experimental data. This is no longer the case for the doped PM metal state ($n = 0.88$), as follows from Fig.~\ref{fig:lsco}(c-d), where only SGA+$1/N$ results reproduce the trends in a semiquantitative manner. For reference, in Fig.~\ref{fig:lsco}(e), we display also SGA+$1/N$ results for the commensurate AF state at slightly lower temperature (cf. Table~\ref{tab:parameters}). The dashed lines in (a-e) are physical real parts of the quasiparticle pole, $\omega_p$, extracted from the computed susceptibilities using the damped-harmonic-oscillator model (cf. Appendix~\ref{appendix:fitting}).  The latter is controlled by bare frequency, $\omega_0$, and damping, $\gamma$ ($\omega_p \equiv \sqrt{\omega_0^2 - \gamma^2}$ for $\omega_0 > \gamma$, otherwise $\omega_p = 0$ and the paramagnon is \emph{overdamped}). Experimental points represent the same propagation frequency, $\omega_p$. Note that, within RPA, paramagnons are overdamped already for $n = 0.88$ along all directions, whereas SGA+$1/N$ result provides propagating ($\omega_p > 0$), but damped spin excitations along $\Gamma$-$X$ direction and the contour following magnetic-zone boundary (cf. the middle panels), in agreement with experiment. The latter results constitute the crucial difference with those of RPA, cf. the explicit comparison of calculated energies $\omega_p$ with experimental data along the $\Gamma$-$X$ line, provided in Fig.~\ref{fig:lsco}(f-g). The PM SGA+$1/N$ calculation slightly overestimates damping along the nodal line [cf. right panel in (c)]. Experimentally, however, paramagnons at $n=0.88$ are close to the overdamping along $\Gamma$-$M$ direction and definitely lose propagating-mode characteristics at $n=0.84$.\cite{Robarts_arXiV_2019}  Full quantitative agreement with RIXS is obtained for the doped AF SGA+$1/N$ case (e). Such a residual magnetic order might be indeed present, since the measurements have been performed at the verge of the stripe phase.\cite{YamadaPhysRevB1998}  In brief, magnetism turns out not to be the crucial factor for the overall spectrum shape, except for the $\Gamma$-$M$ line, where it extends the stability region of magnetic modes. We emphasize that the same Hamiltonian parameters were used in SGA+$1/N$ calculations, both for the undoped ($n=1$) and doped ($n=0.88$) systems (cf. Table~\ref{tab:parameters}). No direct doping-dependence of either the hopping or the Coulomb parameters is necessary to reproduce the data in those two cases. Parenthetically, this may also suggest that $d_{x^2-y^2}$-$d_{z^2}$ hybridization is less relevant to magnetic excitations than previously claimed in terms of the effective Heisenberg-model.\cite{IvashkoPhysRevB2017}  

\begin{table}
  \caption{Summary of model parameters taken to fit the data for $\mathrm{La_{2-\mathit{x}}Sr_\mathit{x}CuO_4}$ (LSCO) and $\mathrm{(Bi,Pb)_2(Sr,La)_2CuO_{6+\delta}}$ (Bi2201). In each case, the next-nearest neighbor hopping is set to $t' = 0.25 |t|$. Phases are marked as: AF -- antiferromagnetic, and PM -- paramagnetic; $n$ is electron concentration.}
\begin{tabular}{ccccccc}
  \hline\hline
  System & Method & $n$ & Phase & $|t|$ (eV) & $U$ (eV) & $k_B T$ (eV) \\
  \hline
  LSCO & SGA+$1/N$ & 1.0 & AF & 0.34 & 2.38 & 0.1224 \\
  LSCO & SGA+$1/N$ & 0.88 & PM & 0.34 & 2.38 & 0.1224 \\
  LSCO & SGA+$1/N$ & 0.88 & AF & 0.34 & 2.38 & 0.1122 \\
  LSCO  & RPA &  1.0 & AF & 0.34 & 2.38 & 0.4420 \\
  LSCO  & RPA & 0.88 & PM & 0.34 & 2.38 & 0.5168 \\
  Bi2201 & SGA+$1/N$ & 0.97 & AF & 0.31 & 1.86 & 0.1116 \\
  Bi2201 & SGA+$1/N$ & $< 0.97$ & PM & 0.31 & 1.86 & 0.1116 \\
  \hline\hline
\end{tabular}
  \label{tab:parameters}
\end{table}

\begin{figure}
  \center
  \includegraphics[width=0.97\linewidth]{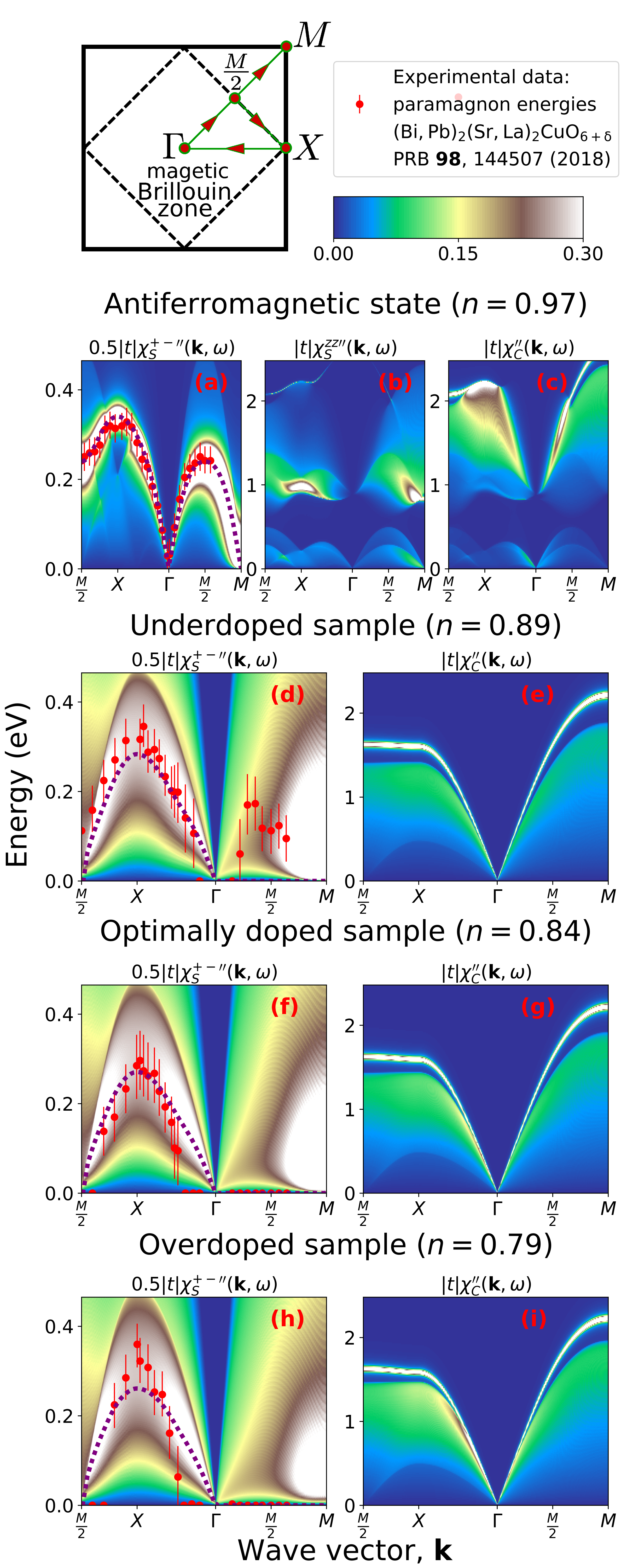}
  \caption{Spin- and charge- dynamic susceptibilities, $\chi_S^{+-}$, $\chi_S^{zz}$, and $\chi_C$ calculated using the present SGA+$1/N$ approach. Red circles correspond to RIXS data for $\mathrm{(Bi, Pb)_2(Sr, La)_2CuO_{6 + \delta}}$.\cite{PengPhysRevB2018}  Note different energy scales for the spin- and charge modes. For computational details see Appendices~\ref{appendix:method} and \ref{appendix:fitting}.  The Brillouin zone contour for panels (a-i) is shown at the top. Dashed lines represent the propagation frequency, $\omega_p$ (cf. Appendix~\ref{appendix:fitting}).}
  \label{fig:bi2201}
\end{figure}

For the sake of completeness, the impact of the local correlations on the charge dynamics is illustrated in Fig.~\ref{fig:splitting}. The color intensity represents imaginary part of the dynamical charge susceptibility $|t| \chi_C^{\prime\prime}(\mathbf{k}, \omega)$ for LSCO at $n=0.88$ (cf. Table~\ref{tab:parameters}), obtained within both SGA+$1/N$ (a) and RPA (b) approximations. Only SGA+$1/N$ yields a clear discrete mode that separates from the continuum. Such a splitting has been also noted within $t$-$J$-$V$ models.\cite{GrecoPhysRevB2016}  Since the charge mode in the cuprates has three-dimensional character \cite{HeptingNature2018} and is sensitive to long-range Coulomb interactions (not included in the Hubbard model), this part of analysis is only qualitative. For reference, we show the measured charge mode energies for $\mathrm{La_{2-\mathit{x}}(Br, Sr)_\mathit{x}CuO_4}$ at $n = 0.875$ (diamonds).\cite{IshiiPhysRevB2017} 

To demonstrate a universal character of the relationship between local correlations and paramagnons,  we provide in Fig.~\ref{fig:bi2201} the same SGA+$1/N$ analysis for the second cuprate family, $\mathrm{(Bi, Pb)_2(Sr, La)_2CuO_{6 + \delta}}$ (Bi2201). In the present situation smaller $U$ and $|t|$ were taken to match the excitation energies in the AF state (cf. Table~\ref{tab:parameters}). Panels~(a-d) exhibit calculated dynamical spin susceptibilities for AF state ($n=0.97$) and PM phase ($n=0.89, 0.84, 0.79$). Dashed lines represent theoretical values of the propagation frequencies and red circles are the RIXS data.\cite{PengPhysRevB2018}  The agreement is quantitative along the $\frac{M}{2}$-$X$-$\Gamma$ contour at all doping levels, from AF insulator ($n=0.97$) to overdoped PM metal ($n=0.79$), except for narrow regions where the paramagnon energies approach to zero. Along the diagonal direction, a strong damping is seen in both experiment and calculations for doped samples, and is marked by the dotted $\omega = \omega_p = 0$ line in Fig.~\ref{fig:bi2201}(d-f). PM SGA+$1/N$ solution yields paramagnon overdamping for $n$ larger by no more than $0.05$ from that observed experimentally. The imaginary part of the longitudinal spin susceptibility ($\chi_S^{zz\prime\prime}$) in the AF state is plotted Fig.~\ref{fig:bi2201}(b). A clear separation of the low-energy and high-energy parts of the e-h continuum is seen in $\chi_S^{zz\prime\prime}$. Close to the $M$-point, the amplitude magnetic mode emerges from the e-h continuum. For the PM case, the corresponding longitudinal part is not displayed as it is equivalent to the transverse susceptibility by the symmetry. Finally, panels (c-i) express calculated imaginary part of the dynamical charge susceptibility. The main qualitative effect of the electronic correlations is detachment of the sharp charge mode from the e-h continuum. This in a non-perturbative effect that does not occur in weak-coupling calculation (cf. Fig.~\ref{fig:splitting}).

\section{Physical discussion}

In correlated electron systems close to localization, Mott physics and magnetic dynamics are well separated and governed by the Hubbard $U$ and kinetic exchange effect on the scale $\ll U$, respectively. The combined SGA+$1/N$ and RPA study (cf. Figs.~\ref{fig:lsco} and \ref{fig:bi2201}) points towards mutual intertwining of these two scales, manifesting itself by persistence of paramagnons in hole-doped systems. Namely, note that the maximal energy of charge excitations in Fig.~\ref{fig:splitting} approaches the value of $U$ (cf. Table~\ref{tab:parameters}), whereas for the spin excitations it is almost \emph{an order of magnitude lower}. Furthermore, superconducting fluctuations are inherently coupled to the spin dynamics, which is exemplified by invoking the identity   $\hat{\mathbf{S}}_i \hat{\mathbf{S}}_j - \frac{1}{4} \hat{n}_i \hat{n}_j \equiv - \hat{B}^\dagger_{ij} \hat{B}_{ij}$,\cite{SpalekPhysRevB1988}  where $\hat{B}^\dagger_{ij} \equiv \frac{1}{\sqrt{2}} \left(\hat{c}^\dagger_{i\uparrow} \hat{c}^\dagger_{j\downarrow} - \hat{c}^\dagger_{i\downarrow} \hat{c}^\dagger_{j\uparrow}\right)$ is the spin-singlet creation operator. The saddle-point (SGA \cite{JedrakPhsRevB2011}) solution (and its DE-GWF extension \cite{SpalekPhysRevB2017}) provide stable AF,\cite{AbramJPhysCondensMatter2017}  charge-density-wave \cite{ZegrodnikPhysRevB2018} or HTSC \cite{SpalekPhysRevB2017} phases, whereas the long-range fluctuations around those states, expressed by the action \eqref{eq:action}, are well accounted for by the extension presented here. In our view, the previous quantitative study of equilibrium and single-particle properties of HTSC,\cite{SpalekPhysRevB2017,FidrysiakJPhysCondensMatter2018,ZegrodnikPhysRevB2018}  combined with the systematic generalization to the collective-quantum-excitation description, provide a direct evidence for a decisive role of strong correlations in high-$T_c$ cuprates. Our analysis can be extended to the $t$-$J$-$U$ or three-band HTSC models. A detailed comparison between our results obtained for a macroscopic system ($> 10^5$ sites) with the results obtained within, e.g., determinant quantum Monte-Carlo method for finite clusters \cite{PengPhysRevB2018} should be also carried out.

\section*{Acknowledgment}

This  work  was  supported  by  Grant  OPUS  No.  UMO-2018/29/B/ST3/02646   from   Narodowe   Centrum   Nauki (NCN).

\appendix

\section{Statistically-consistent-Gutzwiller (SGA) + $\boldsymbol{1/N}$ method}
\label{appendix:method}

Here we provide some of the details of the SGA+$1/N$ technique that we have developed to study magnetic and charge fluctuations in correlated lattice systems. 

The general Hamiltonian that we consider is of the $t$-$J$-$U$-$V$ form

\begin{align}
  \label{eq:tju_model}
  \hat{\mathcal{H}} = & \sum\limits_{ij\alpha\beta\sigma} t_{ij}^{\alpha\beta} \hat{c}^{\alpha\dagger}_{i\sigma} \hat{c}^{\beta}_{j\sigma} + \sum\limits_{i\alpha} U^\alpha \hat{n}_{i\uparrow}^\alpha \hat{n}_{i\downarrow}^\alpha + \sum \limits_{ij\alpha\beta} {}^{'} J_{ij}^{\alpha\beta} \hat{\mathbf{S}}_i^\alpha \hat{\mathbf{S}}_j^\beta + \nonumber \\&+ \sum \limits_{ij\alpha\beta} {}^{'} V_{ij}^{\alpha\beta} \hat{n}_i^\alpha \hat{n}_j^\beta,
\end{align}

\noindent
where $i, j$ and $\alpha, \beta$ denote lattice and orbital indices, respectively. Primes indicate summations over unique pairs of distinct indices. The consecutive terms denote hopping, Hubbard-$U$, intersite Coulomb repulsion, and exchange interactions. Note that, even though here we restrict ourselves to the single-band case, the orbital index is necessary due to the cell-doubling in the antiferromagnetic state.

The starting point is the statistically-consistent approach \cite{JedrakPhsRevB2011,SpalekPhysRevB2017,KaczmarczykPhysRevB2013,KaczmarczykNewJPhys2014} which is based on the variational energy functional $E_G \equiv \langle\Psi_G| \mathcal{H} |\Psi_G\rangle / \langle\Psi_G|\Psi_G\rangle$ with the variational state $|\Psi_G\rangle \equiv \hat{P}_G |\Psi_0\rangle$. Here $|\Psi_0\rangle$ is ``uncorrelated'' wave function (Slater determinant), $\hat{P}_G \equiv \prod_{i\alpha} \hat{P}_{Gi}^\alpha$ with

\begin{align}
\hat{P}_{Gi}^\alpha \equiv & (\lambda^\alpha_{i0} |0\rangle_i{}_i\langle 0| + \lambda^\alpha_{i\uparrow\downarrow} |\uparrow\rangle_i{}_i\langle \downarrow| + \lambda^\alpha_{i\downarrow\uparrow} |\downarrow\rangle_i{}_i\langle \uparrow| +\nonumber \\ &  + \lambda^\alpha_{i\uparrow\uparrow} |\uparrow\rangle_i{}_i\langle \uparrow| + \lambda^\alpha_{i\downarrow\downarrow} |\downarrow\rangle_i{}_i\langle \downarrow| + \lambda^\alpha_{id} |\uparrow\downarrow\rangle_i{}_i\langle \uparrow\downarrow| )
\end{align}

\noindent
is the correlator (controlled locally by six $\lambda$-parameters) that adjusts local many-body configurations in the variational wave function in response to interactions, and $P_\gamma \equiv \langle\Psi_0| \hat{P}_\gamma |\Psi_0\rangle$ are uncorrelated expectation values of the bilinears $\hat{P}_\gamma \equiv \hat{c}^{\alpha\dagger}_{i\sigma} \hat{c}^{\alpha^\prime}_{j\sigma'}$, where $\gamma = (ij\alpha\alpha^\prime\sigma\sigma^\prime)$ is the joint spin-lattice-orbital index. We require that $\hat{P}_G$ is Hermitian which implies that $\lambda^\alpha_{i0}$, $\lambda^\alpha_{i\sigma\sigma}$, and $\lambda^\alpha_{id}$ are real, whereas $\lambda^\alpha_{i\uparrow\downarrow} = \lambda^{\alpha*}_{i\downarrow\uparrow}$. In the absence of spin-orbit coupling, the off-diagonal terms may be safely set to zero as long as the variational ground-state is considered. However, here we will discuss fluctuations that, in general, induce a non-zero values of $\lambda^\alpha_{i\uparrow\downarrow}$ and $\lambda^\alpha_{i\downarrow\uparrow}$.

The statistically-consistent-Gutzwiller free energy functional reads
      
\begin{align}
  \label{eq:free_energy_sga}
      \mathcal{F}_\mathrm{SGA} = -\frac{1}{\beta} & \ln \mathrm{Tr} \exp\Big[-\beta E_G(\mathbf{P}, \boldsymbol{\lambda}) + \nonumber \\ & \beta \boldsymbol{\xi}^\dagger (\mathbf{P} -\hat{\mathbf{P}}) + \beta \mu (\hat{N}_e - N_e)\Big],
    \end{align}

    \noindent
where we have used bold-symbol vector-notation to write down internal-index summations in a compact manner and $\beta \equiv 1/k_BT$ is the inverse temperature. This expression is next optimized with respect to bilinear expectation values $\mathbf{P}$ and correlator parameters $\lambda_\alpha$ under the constraint of fixed particle number.  Additionally, a constraint $\mathbf{P}  = \langle \hat{\mathbf{P}} \rangle$ is enforced by additional set of Lagrange multipliers $\boldsymbol{\xi}$. This step assures that the variational expectation values coincide with those evaluated using the Bogoliubov-de Gennes self-consistent equations. Finally, five more constraints are added for the correlator parameters, $\langle (\hat{P}_{Gi}^{\alpha})^2 \hat{c}^{\alpha\dagger}_{i\sigma} \hat{c}^\alpha_{i\sigma^\prime} \rangle_0 \equiv \langle \hat{c}^{\alpha\dagger}_{i\sigma} \hat{c}^\alpha_{i\sigma^\prime} \rangle_0$, and $\langle (\hat{P}_{Gi}^{\alpha})^2 \rangle_0 \equiv 1$. Note that this method, contrary to the usual variational energy optimization, is applicable also at non-zero temperature. This is essential in the present context as it allows to stay clear of spin-spiral states ubiquitous for Hubbard-type models, both at Hartree-Fock and the saddle-point solution levels at low temperature.\cite{IgoshevJPCM2015} 

The functional \eqref{eq:free_energy_sga} is of saddle-point type and does not capture the long-wavelength collective modes. We now incorporate them by starting from an improved functional

\begin{align}
  \label{eq:sga+1/n_functional}
  \mathcal{F}_{\mathrm{SGA}+1/N} = -\frac{1}{\beta} \ln \int \mathcal{D}[\bar{\boldsymbol{\eta}}, \boldsymbol{\eta}] \mathcal{D} \mathbf{P}_i \mathcal{D} \mathcal{\boldsymbol{\xi}}_i \exp(-\mathcal{S}),
\end{align}

\noindent
with the action

\begin{align}
  \label{eq:action_appendix}
  \mathcal{S} = & \SumInt\limits_i d\tau  \boldsymbol{\eta}_i(\tau)^\dagger (\partial_\tau - \mu) \boldsymbol{\eta}_i(\tau) + \int d\tau E_G(\mathbf{P}_i(\tau), \{x\})  \nonumber\\ & + i \SumInt\limits_{ij} d\tau \boldsymbol{\xi}_{i}(\tau)^\dagger (\hat{\mathbf{P}}_{i}(\tau)  - \mathbf{P}_{i}(\tau)),
\end{align}

\noindent
where $\boldsymbol{\eta}_i(\tau)$ are Grassman fields and the meaning of remaining symbols in Eq.~\eqref{eq:action_appendix} remains the same as in Eq.~\eqref{eq:free_energy_sga}, except for non-trivial imaginary-time dependence. At this point, it is also understood that all the constraints for the correlator parameters have been already solved, so that $\lambda^\alpha_{i\beta} = \lambda^\alpha_{i\beta}(x^\alpha_i)$ with a single variational parameter, $x^\alpha_i$ (there are five constraints for six coefficients per orbital).

Note that the saddle point solution of \eqref{eq:sga+1/n_functional} coincides \emph{exactly} with that obtained from the SGA functional \eqref{eq:free_energy_sga}. The SGA+$1/N$ extension [cf. Eq.~\eqref{eq:sga+1/n_functional}] allows, however, to compute fluctuation corrections to both static and dynamic quantities. Technically, this is done by expanding the energy functional $E_G(\mathbf{P}_i(\tau), \{x\})$ in the time-dependent composite fields $\mathbf{P}_i(\tau)$ to quadratic order and evaluating the resulting free energy $\mathcal{F}_{\mathrm{SGA}+1/N}$ using $1/N$ expansion technique, with $N$ being the number of fermionic flavors. Finally, dynamical spin- and charge susceptibility matrices are evaluated to the leading non-trivial order in $1/N$ (i.e., $N = \infty$), starting from the SGA solution.

\begin{figure*}
  \onecolumngrid
  \centering
  \includegraphics[width=\linewidth]{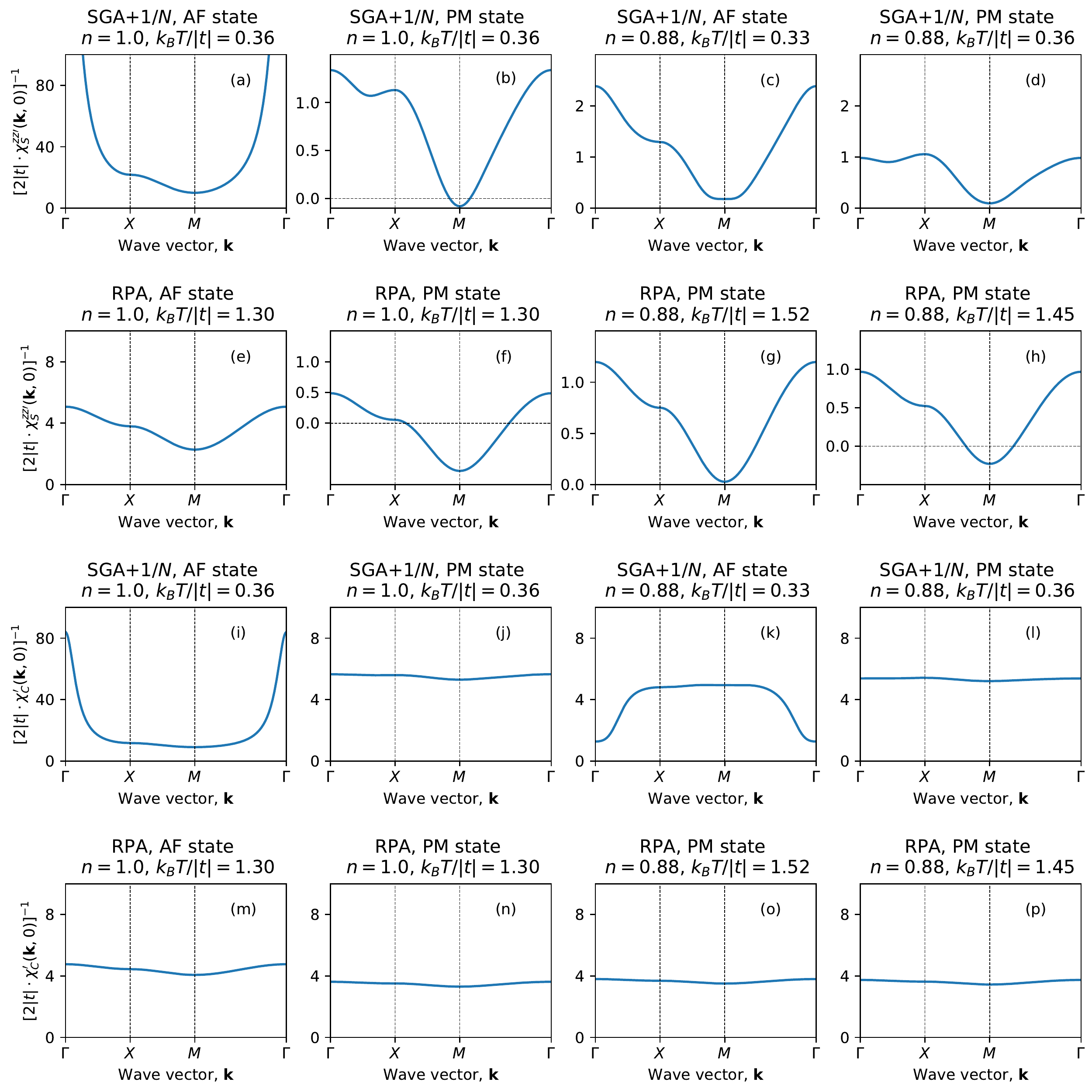}
  \caption{Inverse of the real part of the static ($\omega = 0$) longitudinal spin- [panels (a-h)] and charge- [panels (i-p)] susceptibility for antiferromagnetic (AF) and paramagnetic (PM) states as a function of wave vector. The model parameters are: $t = -0.34$, $t' = 0.25|t|$, $U = 7|t|$.  Calculations were performed using the present SGA+$1/N$ method [panels (a-d) and (i-l)] and RPA [panels (e-h) and (m-p)]. Temperature and doping level are displayed on the top of each panel.}
  \label{fig:stability_lsco}
    \twocolumngrid
\end{figure*}

\section{Numerical details}
\label{appendix:numerical_details}

In this analysis we restrict to the Hubbard model with nearest- and next-nearest-neighbor hoppings, $t < 0$ and $t' = 0.25|t|$, respectively. Aside from tests and benchmarks (encompassing symmetry analysis in external applied Zeeman field and for various microscopic Hamiltonians), the exchange and intersite Coulomb interactions have been set to zero. The model parameters for two considered cuprate compounds, $\mathrm{La_{2-\mathit{x}}Sr_\mathit{x}CuO_4}$ (electron concentration $n=1, 0.88$) and $\mathrm{(Bi,Pb)_2(Sr,La)_2CuO_{6+\delta}}$ ($n=0.97, 0.89, 0.84, 0.79$) are summarized in Table~I of the main text. All calculations have been performed at $T > 0$ for $N \times N$ two-site magnetic cells ($N = 300$; $18 \times 10^4$ sites in total) with imposed periodic boundary conditions. A mesh of 200 frequencies was used to generate the dynamical susceptibility color maps of and a small imaginary part was added to frequency while performing analytic continuation $i \omega_n \rightarrow \omega + i \epsilon$ with $\epsilon = 0.008|t|$. While computing real parts of susceptibility $\epsilon$ was set to zero, but small uniform $\mathbf{k}$-space disorder was introduced to minimize finite-size effects. The target absolute accuracy for dimensionless variational parameters  was set to $10^{-10}$. Algorithmic Hamiltonian block-reduction was implemented at the level of variational calculation and susceptibility-matrix evaluation. Monte Carlo tree-search algorithm (MCTS) implemented in FORM system \cite{FORM_citation} was used for symbolic energy functional optimization. GNU Scientific Library was employed for multidimensional optimization and matrix operations. The computations were performed on a dedicated supercomputer EDABI (Jagiellonian University, Krak\'ow, Poland).

\begin{figure*}
    \onecolumngrid
  \centering
  \includegraphics[width=\linewidth]{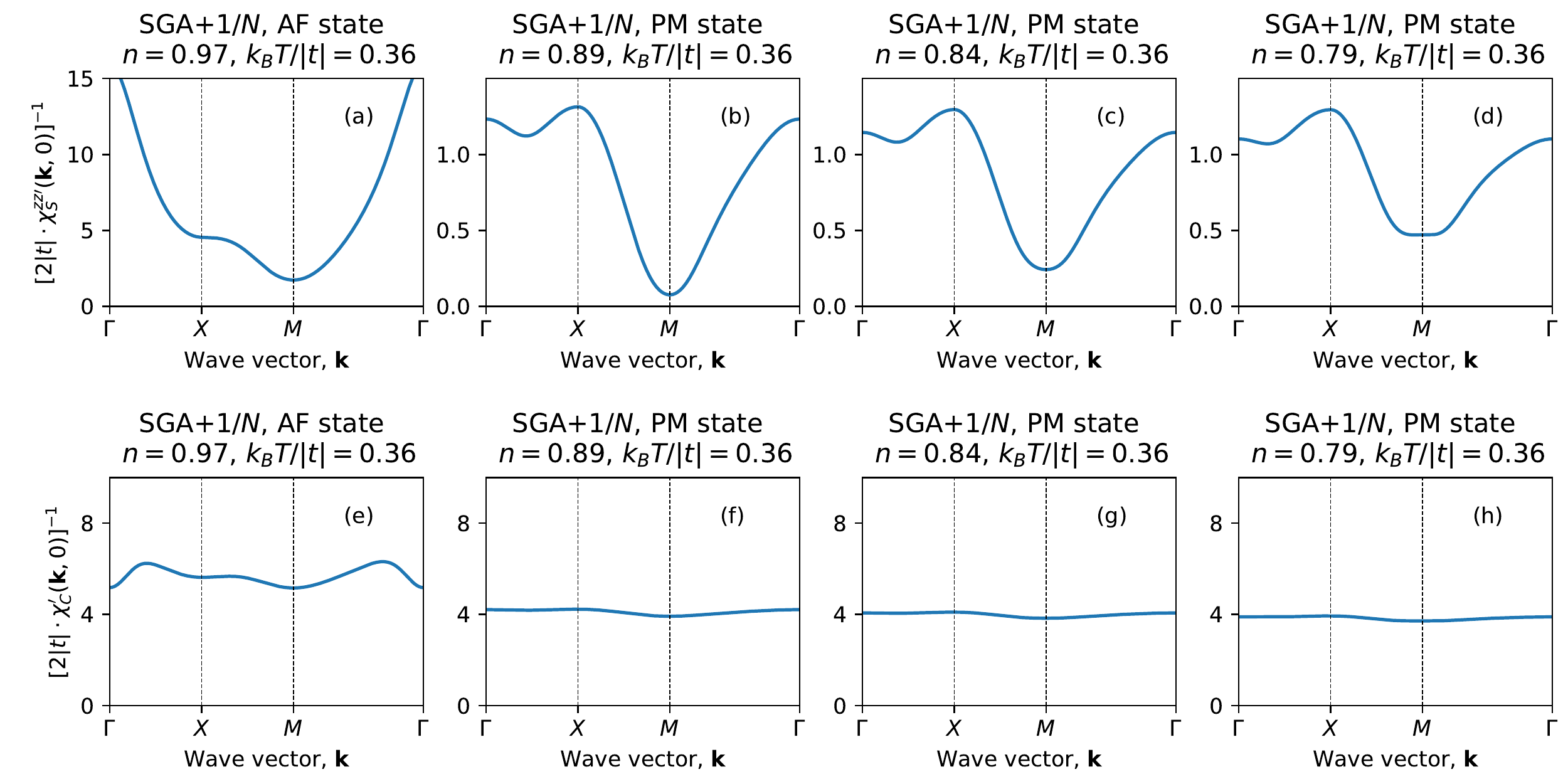}
  \caption{Inverse of the real part of static ($\omega = 0$) longitudinal spin- [panels (a-h)] and charge- [panels (i-p)] susceptibility for antiferromagnetic (AF) and paramagnetic (PM) states as a function of wave vector, calculated using the present SGA+$1/N$ method at various doping levels. The model parameters are: $t = -0.31\,\mathrm{eV}$, $t' = 0.25|t|$, $U = 6|t|$. Temperature is set to $k_B T = 0.36 |t|$ for all the dopings specified.}
  \label{fig:stability_bi2201}
  \twocolumngrid
\end{figure*}

\begin{figure}
  \centering
  \includegraphics[width=\linewidth]{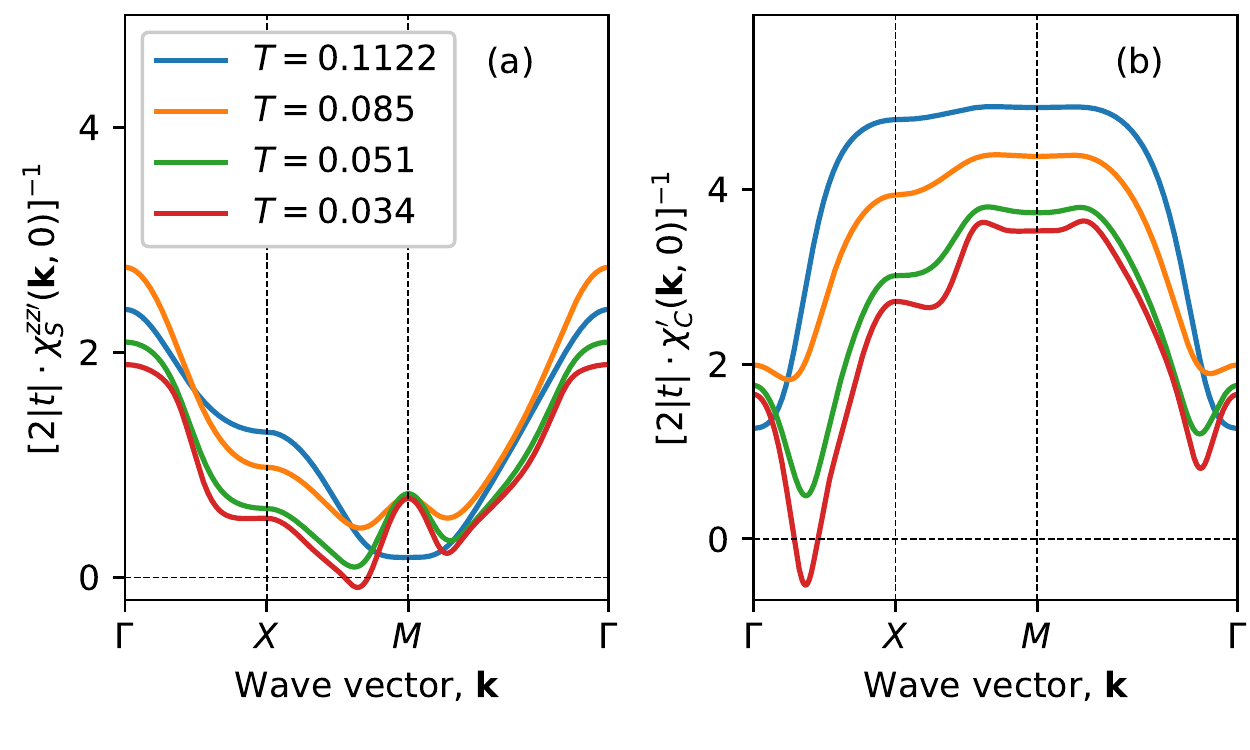}
  \caption{Inverse of the real part of static ($\omega = 0$) longitudinal spin (a) and charge (b) susceptibility for AF state and $n=0.88$, calculated using the present SGA+$1/N$ approach. The model parameters are $t = -0.34\,\mathrm{eV}$, $t' = 0.25|t|$, $U=7|t|$. Temperatures in panel (a) are given in eV. As temperature is lowered, a near simultaneous dynamical incommensurate spin- and charge-instability of the AF state is generated.}
  \label{fig:stability_temp}
\end{figure}

\section{Phase stability}
\label{appendix:phase_stability}

A successful optimization of the variational free-energy functional does not ensure the solution stability in the presence of higher-order effects, such as fluctuation corrections. Here we demonstrate that, for the selection of temperatures provided in Table~I of the main text,  all the considered phases are stable against spin and charge fluctuations. In Fig.~\ref{fig:stability_lsco}(a-h) we plot inverse real part of the zero-frequency dynamical longitudinal spin susceptibility for the parameters employed for $\mathrm{La_{2-\mathit{x}}Sr_\mathit{x}CuO_4}$ ($t = -0.34$, $t' = 0.25|t|$, $U/|t| = 7$). The undoped system ($n=1$) exhibits stable commensurate AF state within the SGA+$1/N$ approximation for $k_B T = 0.36 |t| \ll |t|$ [cf. (a)], whereas paramagnetic state is unstable against magnetic fluctuations (b). For the same temperature, SGA+$1/N$ yields stable PM state for doped ($n=0.88$) system [panel (c)] and commensurate AF state for slightly lowered $k_B T = 0.33 |t|$ (d). To obtain a similar behavior within RPA approximation, one needs to break the hierarchy of energies $k_B T \ll |t| \ll U$, as demonstrated in panels (e-h). In particular, even for $k_B T = 1.45 |t| > |t|$ the system is magnetically unstable at $n=0.88$ (h), which renders RPA inadequate in the strong-coupling regime. Similar analysis for charge susceptibilities, summarized in panels (i-p), shows that there are no dynamical instabilities to charge ordered phases for the selected parameter values.

We have also performed a similar analysis for the parameters suitable for $\mathrm{(Bi,Pb)_2(Sr,La)_2CuO_{6+\delta}}$ (Bi2201). The results are  summarized in Fig.~\ref{fig:stability_bi2201}, proving the dynamical stability of the SGA+$1/N$ solution against spin and charge fluctuations for all considered doping levels.

Finally, it is instructive to observe how the extensively studied \cite{IgoshevJPCM2015} incommensurate phases emerge as the temperature is lowered. For $n=0.88$, starting from the PM phase, a commensurate AF state appears around $k_B T = 0.33 |t|$. A further cooling generates near instantaneous spin- and charge- dynamical instability of the AF order, as demonstrated in Fig.~\ref{fig:stability_temp}. By proper selection of temperature, one can thus stay clear of incommensurate magnetic states.

\section{Damped harmonic oscillator modeling of the SGA+$\boldsymbol{1/N}$ and RPA results: Details of fitting the data}
\label{appendix:fitting}

\begin{figure}
  \centering
  \includegraphics[width=\linewidth]{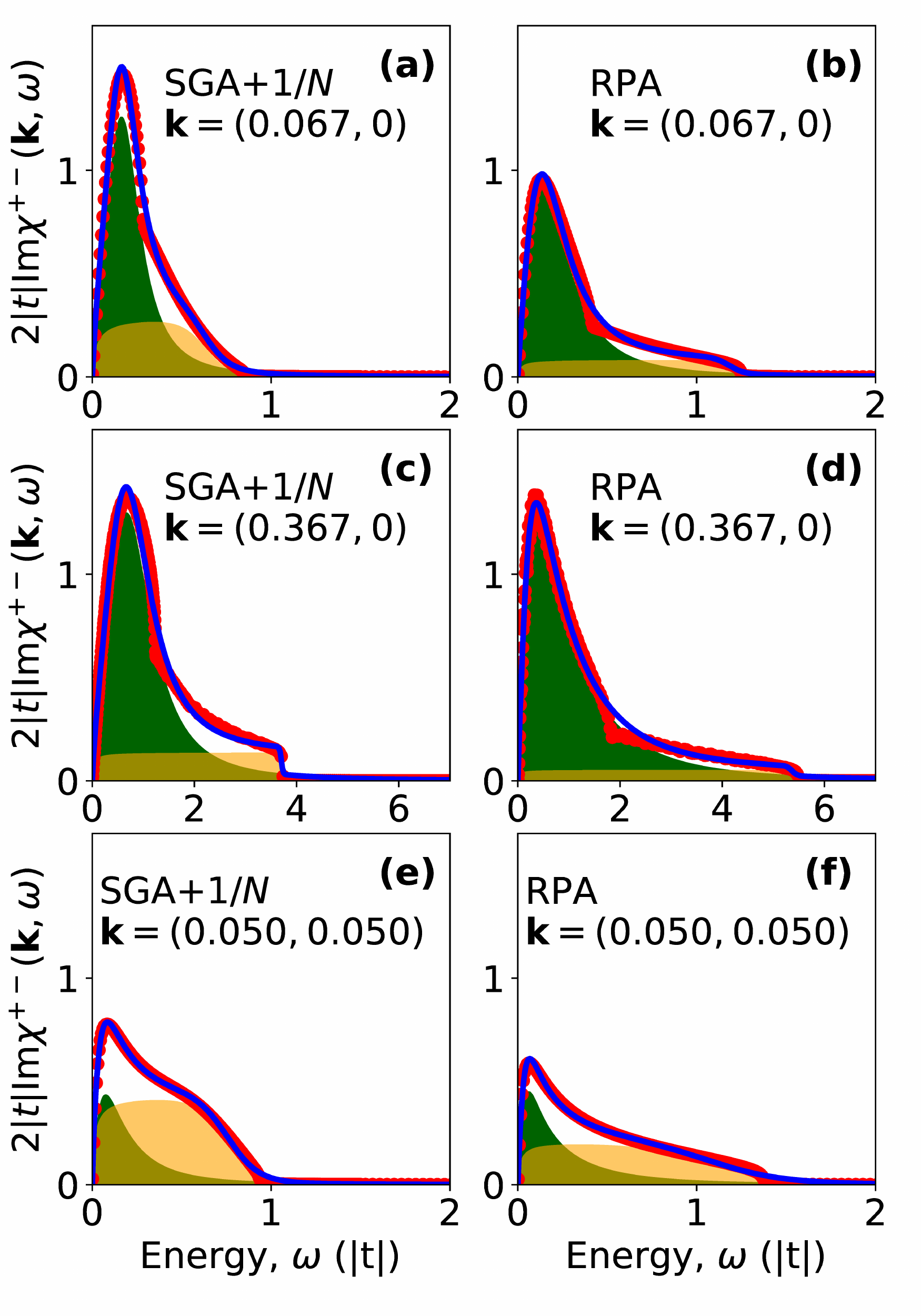}
  \caption{Exemplary fits of the total dissipative part of susceptibility $\chi^{\prime\prime}_\mathrm{total} \equiv \chi^{\prime\prime}_\mathrm{oscillator} + \chi^{\prime\prime}_\mathrm{incoh}$ (blue solid curve) to the imaginary part of dynamical susceptibility $\mathrm{Im} \chi^{+-} (\mathbf{k}, \omega)$  calculated using SGA+$1/N$ [panels (a), (c), (e)] and RPA [panels (b), (d), (f)] (red circles) for doping $n = 0.88$. The wave vectors, $\mathbf{k}$, in the panels are given in the units of $2\pi/a$ with $a$ being lattice spacing. The damped-oscillator and background partial contributions are marked in green and orange, respectively.}
  \label{fig:fit_n088}
\end{figure}

\begin{figure}
  \centering
  \includegraphics[width=\linewidth]{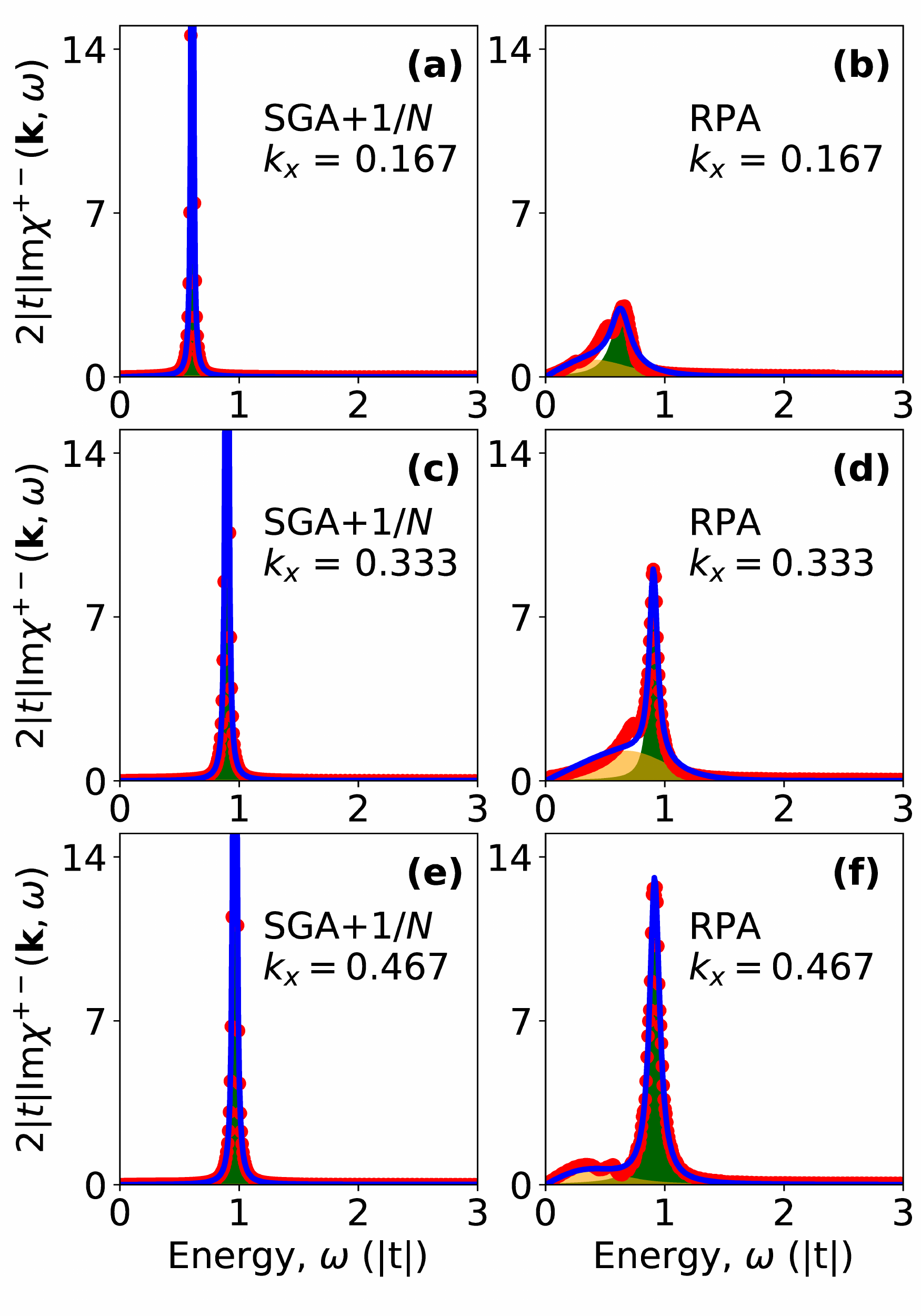}
  \caption{Same as in Fig.~\ref{fig:fit_n088}, but at half-filling ($n = 1$). The sharp peaks represent a true magnon excitation in the stable AF phase.}
  \label{fig:fit_n1}
\end{figure}

Whereas the principal result of the present contribution is evaluation of the magnetic response functions  for strongly-correlated states (cf. color maps Figs.~1 and 3 of main text), comparison of obtained theoretical results with available experimental data requires further processing. In this section, we describe this secondary analysis.

The experimental RIXS spectra are typically modeled by the dissipative part of damped harmonic-oscillator response of the form

\begin{align}
  \label{eq:damped_oscillator}
  \chi^{\prime\prime}_\mathrm{oscillator}(\mathbf{k}, \omega) = \frac{2 F(\mathbf{k}) \gamma(\mathbf{k}) \omega}{\left[\omega^2 - \omega_0^2(\mathbf{k})\right]^2 + 4 \gamma(\mathbf{k})^2 \omega^2},
\end{align}

\noindent
with the particle-hole and multi-magnon background added extra. Here $\omega_0(\mathbf{k})$ and $\gamma(\mathbf{k})$ are wave-vector-dependent bare frequency and damping coefficients, and $F(\mathbf{k})$ is scaling factor. Equation~(\ref{eq:damped_oscillator}) describes a \emph{damped propagating} mode (paramagnon) of energy $\omega_p(\mathbf{k}) \equiv \sqrt{\omega_0^2(\mathbf{k}) - \gamma^2(\mathbf{k})}$ if $\omega_0 > \gamma$. Otherwise, for $\gamma > \omega_0$, the quasiparticle pole becomes purely imaginary and the mode is \emph{overdamped}. Apart from $\omega_0$ and $\omega_p$, one can also define the third frequency, $\omega_\mathrm{max}$, for which expression~(\ref{eq:damped_oscillator}) attains maximum. Physically, the most relevant one, and also reported in many of the experimental works, is $\omega_p$ that we also provide in the present study.

We introduce particle-hole background of the form

\begin{align}
  \label{eq:continuum_function}
  \chi^{\prime\prime}_\mathrm{incoh} \equiv \frac{A(\mathbf{k}) \omega}{1 + B(\mathbf{k}) \omega} n_F[C(\mathbf{k})(\omega - D(\mathbf{k}))],
\end{align}

\noindent
where $n_F(x) \equiv (\exp(x) + 1)^{-1}$ is the Fermi function, and $A, B, C, D$ are positive $\mathbf{k}$-dependent coefficients. This function describes a featureless continuum that is softly cut off on the low-energy side and suppressed above threshold $\omega \sim D$. We have found that the function~(\ref{eq:continuum_function}) is flexible enough to accurately model the particle-hole background in the wide doping range. Note that none of the employed approximations accounts for the weak multi-magnon peak, seen in some RIXS experiments.\cite{Robarts_arXiV_2019}  Hence, there is no need to subtract this feature.

In Fig.~\ref{fig:fit_n088} we show exemplary fits of the total dissipative part of susceptibility $\chi^{\prime\prime}_\mathrm{total} \equiv \chi^{\prime\prime}_\mathrm{oscillator} + \chi^{\prime\prime}_\mathrm{incoh}$ (blue solid curve) to the imaginary part of dynamical susceptibility $\chi^{+-\prime\prime} (\mathbf{k}, \omega)$, calculated using the present SGA+$1/N$ (left panels) and RPA (right panels) approximations. The total response is decomposed into a step-like particle-hole continuum (orange color) and the peak that is modeled by a damped harmonic oscillator (displayed in green). From Fig.~\ref{fig:fit_n088}(a-f) it is apparent that \emph{the main effect of electronic correlations is to compress the incoherent part of the spectrum and reduce the paramagnon damping} (these correlations are incorporated only into the SGA+$1/N$ calculation). However, the magnetic peak maximum $\omega_\mathrm{max}$ is actually shifted to higher energies due to local Hubbard-interaction physics [as seen particularly well in panels (c) and (d)]. This effect is counterintuitive as it opposes the behavior of the incoherent excitations that are transferred to lower energy, but it is necessary to match the experimental data for the cuprates. Similar fits, performed at half-filling ($n = 1$) are shown in Fig.~\ref{fig:fit_n1}. In this case the sharp coherent peak is a true magnon excitation and a small residual background represents residual particle-hole excitations, present for $T > 0$ and in the situation with not too strong correlations ($U/|t| \sim 7$).

\bibliography{bibliography}
\end{document}